\documentclass[twocolumn,amsmath,amssymb]{revtex4}

\usepackage{graphicx}

\begin{document}

\title{A simple and robust method for connecting small-molecule drugs
using gene-expression signatures}
\author{Shu-Dong Zhang}
\email{sdz1@le.ac.uk}
\author{Timothy W. Gant}

\affiliation{MRC Toxicology Unit, Hodgkin Building, Lancaster
Road, University of Leicester, Leicester, UK}


\begin{abstract}

Interaction of a drug or chemical with a biological system can
result in a gene-expression profile or signature characteristic of
the event. Using a suitably robust algorithm these signatures can
potentially be used to connect molecules with similar
pharmacological or toxicological properties. The Connectivity Map
was a novel concept and innovative tool first introduced by Lamb
et al to connect small molecules, genes, and diseases using
genomic signatures [Lamb et al (2006), Science 313, 1929-1935].
However, the Connectivity Map had some limitations,
particularly there was no effective safeguard against false
connections if the observed connections were considered on an
individual-by-individual basis. Further when several connections
to the same small-molecule compound were viewed as a set, the
implicit null hypothesis tested was not the most relevant one for
the discovery of real connections.
Here we propose a simple and robust method for constructing the
reference gene-expression profiles and a new connection scoring
scheme, which importantly allows the valuation of statistical
significance of all the connections observed. We tested the new
method with the two example gene-signatures (HDAC inhibitors and
Estrogens) used by Lamb et al and also a new gene signature of
immunosuppressive drugs. Our testing with this new method shows
that it achieves a higher level of specificity and sensitivity
than the original method. For example, our method successfully
identified raloxifene and tamoxifen as having significant
anti-estrogen effects, while Lamb et al's Connectivity Map failed to
identify these. With these properties our new method has potential
use in drug development for the recognition of pharmacological and
toxicological properties in new drug candidates.

\end{abstract}


\maketitle

\section{Introduction}

One of the most fundamental challenges in all forms of  'omic
technologies is the connection of biological event signatures with
others previously derived to allow the recognition of new molecule
properties or biological alteration. Simple, robust, and efficient
matching methods are required to connect a new gene signature with
those in a database. This problem was first tackled by Lamb et al
 \citep{Lamb-etal-Science-2006} who introduced the
Connectivity Map as a resource and tool to connect small-molecule
drugs, genes, and diseases. Lamb et al's Connectivity Map achieved a
good degree of success, but also suffered from several
deficiencies, particularly an inability to apply a measure of
statistical validity at the individual reference signature level
to allow rational filtering of the results to exclude false
connections. We took the method of Lamb as a basis for development
and have derived a simple, robust and statistically testable
method for making connections between biological event signatures.
The method was tested with genomic signatures resulting from small
molecule interactions in cells, but also could be applied to any
form of signature such as those from proteomic or metabolomic
science.

The main assumption behind the concept of Connectivity Map is that
a biological state, whether physiological, pathological, or
induced with chemical or genomic perturbations, can be described
in terms of a genomic signature, eg., the genome-wide mRNA levels
as measured by DNA microarray technologies. The working of the
Connectivity Map involves several key components. First, a large
collection of pre-built reference gene-expression profiles serve
as the core database, where each reference profile characterizes a
well-defined biological state. Secondly, a query gene signature
from some specific studies. A query gene signature is basically a
short (as compared to the list of genes in a typical reference
profile) list of genes most relevant and important to characterize
the biological state of the researchers' interest. Finally, a
pattern matching algorithm or similarity metric defined between a
query gene signature and a reference gene-expression profile to
quantify the closeness or connection between the two biological
states. The Connectivity Map could be used by biomedical
researchers to find connections between the reference biological
states and those of their own interest. This may lead to new
findings of unexpected connections and testable new biological
hypotheses. In this paper, we present a new framework for the
construction of reference profiles and new connection scoring
scheme and testing procedures for the observed connections. We
compare our method with that of Lamb et al, and show that more
robust results are achieved using our method. In particular our method not only offers more principled statistical procedures for testing connections, but more importantly it provides effective safeguard against false connections while at the same time achieves much increased sensitivity. As a consequence it can benefit the end users by saving them enormous amount of time and resources in pursuing new biological hypotheses based on the findings of connectivity maps.

\section{Methods}

\subsection{Construction of reference profiles}

For the first-generation connectivity map, Lamb et al carried out
a series of gene-expression profiling experiments
\citep{Lamb-etal-Science-2006}, using $164$ distinct
small-molecule compounds in a few selected human cell lines. Each
treatment instance consisted of one treatment sample and one (or
more) vehicle control samples, whose genome wide mRNA levels were
measured using Affymetrix GeneChip microarrays. In total $564$
samples were microarrayed, which represented $453$ different
treatment instances. For example, treatment instance ID988
consisted of 1 treatment sample and 6 vehicle control samples. The
treatment sample was obtained by treating human MCF7 cells with
$100 nM$ estradiol for 6 hours. The control samples were obtained
by treating MCF7 cells with vehicle control for 6 hours.
A gene-expression profile was constructed for each treatment
instance, in which the relative expression (treatment relative to
the control) of all measured genes were specified, and sorted in
descending order.
A query gene signature, obtained and ordered in the same manner,
can be compared to each reference profile in the Connectivity Map to
calculate a connectivity score. For ``{\em positive
connectivity}'', the up-regulated genes of the query signature
find matches near the top of the reference profile, and the
down-regulated genes find matches near the bottom of the reference
profile. For ``{\em negative connectivity}'', the matches are
opposite.

We obtained Lamb et al's data set (Accession Number GSE5258) from
the GEO (Gene Expression Omnibus) database, and rebuilt $453$
reference gene-expression profiles using a new ranking scheme
based on the following guiding principles: (1) A treatment
instance was defined relative to a control, thus the effect of the
treatment could be characterized by the relative differential
expression status of all the genes together, (2) different genes
were affected to different extents by the treatment, so genes
which showed a greater differential expression should have more
weight in characterizing this treatment, and (3) up- and
down-regulated genes should be treated equally in a unified
manner. This meant that a 2-fold down-regulated gene was considered as
equally important as a 2-fold up-regulated gene in defining a
reference profile. There are several reasons for the choice of treating up- and down-regulated genes equally. Theoretically, unless we have a lot of further information about so many genes on the microarray it is difficult to decide whether this 2-fold up-regulated gene is more important than that 2-fold down-regulated gene or the opposite. So a logical choice is to assign them equal weights. Another reason is the consideration of symmetry: if a gene is 2-fold up-regulated in sample 1 versus sample 2, it can also be viewed as 2-fold down-regulated in sample 2 versus sample 1. We should emphasize that assigning two genes equal weights does not imply in any sense they share the same molecular mechanism.  Even two up-regulated genes with the same fold changes could have very different molecular mechanisms as to why they increased their expression. To adhere to the above guidelines, an obvious
choice for organizing the genes is the logarithm of the expression
ratio (treatment over control). Thus instead of treating the down-
and up-regulated genes separately as in the methods of Lamb et al,
we ordered genes in a reference profile by the absolute value of
their expression log-ratios. Therefore the most differentially
expressed genes (either up or down) appear first in the list, and
those non-differentially expressed genes appear at the bottom of
the list. In this way, the genes are ordered by their importance
in defining the reference profile. It is then straightforward to
assign ranks to them. Suppose there are in total $N$ genes, the
first gene in the list will be assigned a rank $N$ if it is
up-regulated, or a rank $-N$ if it is down-regulated. In general
the $i$th gene in the list will be ranked with $(N-i+1)$ for
up-regulation or $-(N-i+1)$ for down-regulation. With this new
ranking method, the importance of a gene is reflected by the
absolute value of its rank, while the sign of its rank indicates
its regulation status. The consequence and advantage of this
method for creating reference profiles is that attaching
statistical significance to the connection observed is a
relatively straightforward step.

\subsection{The new scoring scheme} \label{sec-scoring}

A query gene signature
can be an ordered gene list, or just a collection of genes without
specific ordering among them. We will refer to these two types of
query gene signatures as {\em ordered} and {\em unordered} gene
signature respectively. For an ordered gene signature, we rank the
genes in the list in the same way as for a reference profile.
Namely, the most important (differentially expressed) gene in the
signature will be assigned a rank $m$ or $-m$ depending on its
regulation status, where $m$ is the number of genes in the
signature. While the least important gene in the signature be
ranked $1$ or $-1$.

Let $\mathbf{R}$ denote a reference gene-expression profile, and
$\mathbf{s}$ a query gene signature. We define the {\em connection
strength} between $\mathbf{R}$ and $\mathbf{s}$ as
\begin{equation}
C(\mathbf{R},\mathbf{s})=\sum_{i=1}^{m}R(g_i)s(g_i),
\label{eq-C(R,s)}
\end{equation}
where $g_i$ represents the $i$th gene in the signature, $s(g_i)$
is its signed rank in the signature, and $R(g_i)$ is this gene's
signed rank in the reference profile. It is worth noting some
properties of the connection strength defined above: (1) if a gene
has the same regulation status (either up- or down-regulation) in
both the reference and the query, it will make a positive
contribution to the connection strength, otherwise its
contribution will be negative; (2) the magnitude of a gene's
contribution to the connection strength is determined by its
position in both lists; and (3) a gene signature with some of its
genes contributing positively and others negatively will have an
overall low connection strength, because the positive and negative
contributions cancel each other to some extent. Therefore
calculating the connection strength between a gene signature and a
reference profile, the maximum connection strength achievable is
the situation where the $m$ genes in the signature match the first
$m$ genes in the reference profile in the correct order, and their
regulation status also match. In such a case, the maximum positive
connection strength is, for an {\em ordered} gene signature,
\begin{eqnarray}
C^{o}_{max}(N,m)=\sum_{i=1}^{m}(N-i+1)(m-i+1).
\label{eq-C_max-ordered}
\end{eqnarray}
In another equally interesting situation, where the $m$ genes in
the signature match the first $m$ genes in the reference profile
in the right order, but the sign of each gene in the query is
different from its sign in the reference, the connection strength
is $-C^{o}_{max}(N,m)$, the opposite of
Eq.(\ref{eq-C_max-ordered}).

For an unordered query signature, all the genes in the list have
equal weight because there is no particular ordering among them.
The calculation of connection strength is the same as
Eq.(\ref{eq-C(R,s)}), the only difference being that $s(g_i)=1$ if
gene $g_i$ is up-regulated, or $s(g_i)=-1$ if it is
down-regulated. Consequently, the maximum magnitude of connection
strength for an {\em unordered} signature is
\begin{eqnarray}
C^{u}_{max}(N,m)=\sum_{i=1}^{m}(N-i+1). \label{eq-C_max-unordered}
\end{eqnarray}

Given a query signature gene and a reference gene-expression
profile, we can calculate their connection score by
\begin{eqnarray}
c=C(\mathbf{R},\mathbf{s})/C_{max}(N,m). \label{eq-c}
\end{eqnarray}
So a connection score $c=1$ means that the gene signature has the
maximum positive connection strength with the reference profile,
which indicates that the experimental condition that gave rise to
this gene signature had the strongest possible correlation with
the treatment instance that generated the reference profile. A
connection score $c=-1$ indicates that the two experimental
perturbations were most inversely correlated. In general, a
connection score $c$ will be within the range of $(-1, 1)$.

\subsection{Connection Testing} \label{sec-testing}

As for most biomedical experiments with unavoidable biological and
technical variation, statistical significance is a crucial aid to
the interpretation and subsequent validation of the result. Here
we propose calculating the p-value associated with a connection
score by testing the following null hypothesis.

{\bf Null hypothesis $H_0$:} For a reference gene-expression
profile $\mathbf{R}$ and a query gene signature $\mathbf{s}$, the
null hypothesis $H_0$ states that there is no underlying
biological connection between the two, and that the query
signature $\mathbf{s}$ is merely a random $m$-gene signature, as
generated by Procedure 1 described below.

{\bf Procedure 1:} Generation of a random $m$-gene signature. Let
$\mathbf{R}$ be a given reference gene-expression profile of $N$
genes. Select $m$ genes sequentially and randomly from the $N$
genes (without replacement), and assign $+1$ (up-regulation) or
$-1$ (down-regulation) randomly with equal probability to each of
the $m$ selected genes. If this gene signature is to be used as an
ordered list, its order is just the order in which the $m$ genes
are selected; or if this gene signature is to be used as merely a
collection of genes, then order is irrelevant.

Given a reference profile $\mathbf{R}$ and a gene signature
$\mathbf{s}$, we calculate their connection score $c$ by
Eq.(\ref{eq-c}), and the two-tailed p-value associated with this
observed connection score is
\begin{eqnarray}
p=\mbox{Prob} \left\{|\tilde{c}|\ge |c|\mid H_0 \right\},
\end{eqnarray}
where $\tilde{c}$ is the connection score between a random gene
signature and the reference profile. To estimate the p-value, a
large number (eg., $10^5$) of random connection scores can be
generated using Procedure 1 and Eq.(\ref{eq-c}), the proportion of
random scores that are no less than the observed scores $c$ in
absolute value is an estimate of the two-tailed p-value.

The $453$ individual treatment instances of the data set GSE5258
were created using only $164$ distinct small-molecule compounds.
Some treatment instances were replication experiments using the
same compound at the same or different doses. It is thus
interesting to consider all the treatment instances of the same
compound as a set, and to assess the overall connection of the set
with a query gene signature. We define the connection score for a
treatment set as follows
\begin{eqnarray}
t=\frac{1}{n} \sum_{i=1}^{n}c_i, \label{eq-setcscore}
\end{eqnarray}
where $n$ is the number of individual treatment instances
belonging to the treatment set, $c_i$ is the connection score of
the $i$th instance. To test the significance of a treatment set as
a whole. We used the following null hypothesis,

{\bf Null hypothesis $H^{set}_0$:} Where $T$ denotes a set of
treatment instances, $\mathbf{R_i}$ the reference profile based on
treatment instance $i$, and $\mathbf{s}$ a query gene signature,
the null hypothesis $H^{set}_0$ states that there is no underlying
biological connection between the gene signature $\mathbf{s}$ and
any of the reference profiles in T. The query signature is merely
a random gene list generated by Procedure 1.

Thus the null hypothesis $H_0^{set}$ is an extension of $H_0$ to a
higher level. Alternatively, $H_0$ can be viewed as a special case
of $H^{set}_0$, in which there is only one treatment instance in
the treatment set. Once the connection score for a set is
observed, its associated p value can be estimated in a similar
way: a large number of random gene signatures are generated using
Procedure 1, and the connection score of the set to each of the
random gene signatures is calculated using
Eq.(\ref{eq-setcscore}); the proportion of random connection
scores that are greater than the observed score in absolute value
is an estimation of the p value.

\section{Results}

\begin{table}
\caption{\label{tab-rds01-Lamb}   The connectivity scores of the
random gene signature rds01 to the reference profiles in the Lamb
Connectivity Map.
}

\begin{tabular}{cclccc}
\hline
rank&ID&compound&score&up&down\\
\hline
1&1080&sirolimus&1&0.232&-0.578\\
2&913&colforsin&0.953&0.245&-0.527\\
3&1138&phentolamine&0.912&0.316&-0.423\\
4&1048&alpha-estradiol&0.886&0.324&-0.394\\
5&1115&phenanthridinone&0.869&0.379&-0.325\\
$\cdots$&$\cdots$&$\cdots$&$\cdots$&$\cdots$&$\cdots$\\
112&885&5186223&0.363&0.137&-0.157\\
113&3&metformin&0.342&0.158&-0.119\\
114&663&U0125&0&0.405&0.194\\
115&124&mesalazine&0&0.371&0.256\\
$\cdots$&$\cdots$&$\cdots$&$\cdots$&$\cdots$&$\cdots$\\
369&1008&geldanamycin&0&-0.43&-0.339\\
370&1064&17AAG&0&-0.436&-0.395\\
371&605&monastrol&-0.38&-0.114&0.177\\
372&494&fluphenazine&-0.392&-0.113&0.187\\
$\cdots$&$\cdots$&$\cdots$&$\cdots$&$\cdots$&$\cdots$\\
449&601&MK-886&-0.834&-0.303&0.336\\
450&604&arachidonic acid&-0.855&-0.301&0.354\\
451&387&estradiol&-0.901&-0.162&0.528\\
452&379&cobalt chloride&-0.916&-0.238&0.464\\
453&378&tacrolimus&-1&-0.23&0.536\\
\hline
\end{tabular}
\end{table}

\subsection{Querying the connectivity maps with random gene
signatures}

To compare the specificities of Lamb et al's Connectivity Map and the
connectivity map presented here, we generated random gene
signatures and tested these random gene signatures in both. The
first example was a random gene signature, rds01, which contained
$25$ Affymetrix probe-set IDs randomly selected from the $22283$
IDs on the Affymetrix HG-U133A microarray platform. Querying Lamb et al's Connectivity Map with this signature, we obtained the
connectivity scores of rds01 to all the $453$ individual treatment
instances. The results are shown in Table \ref{tab-rds01-Lamb},
which suggests that this signature has positive connections to
$113$ individual reference profiles, with connectivity scores
ranging from $1$ to $0.342$; and that $83$ individual reference
profiles have negative connections with rds01, with connectivity
scores ranging from $-1$ to $-0.38$; the remaining $257$ reference
profiles have null connections (connectivity score 0). However
rds01 was a random gene signature and so all these positive and
negative connections must be false. We then used this same random
signature to query the connectivity map presented in this paper
(Table \ref{tab-rds01-sdz}). With a p value calculated for each
observed connection score, we can control the expected number of
false connections by setting an appropriate threshold p value. In
this paper, the threshold p value is generally set at $1/N$, where
$N$ is the number of null hypotheses being tested simultaneously.
In this example, $N=453$ (the total number of individual treatment
instances in the database and hence the number of null hypothesis
being tested at the treatment instance level). So a connection
with a p value $p<1/453=0.0022$ is considered as statistically
significant. The setting of the threshold p value at $1/N$ was
intended to control the expected number of false connections at 1,
we thus expected to have one false connection on average among all
the connections declared as significant. Table \ref{tab-rds01-sdz}
shows that our connectivity map gave the correct result, ie., no
significant connection between this random signature and any of
the treatment instances. This compares with the $196$ connections
($113$ positive and $83$ negative) found with the method of Lamb
et al. Note that to control the expected number of false
connections more precisely, the threshold hold p value should be
set at $1/N_0$, where $N_0$ is the number of true null hypothesis.
Of course in a general situation $N_0$ is unknown, so it has to be
estimated, for example, using the methods developed in
 \citep{Storey&Tibshrirani-pnas2003,
Zhang&Gant-Bioinformatics2004}. In this paper, we set threshold p
value at $1/N$ for simplicity. Since $N_0\le N$, our criteria tend
to be slightly conservative, meaning that the actual number of
false connections on average will be $\le 1$.

The second random gene signature, rds02, consisted of 189
Affymetrix probe-set IDs randomly selected from $22283$. The full
results of querying Lamb et al's Connectivity Map with this signature
can be found in the supplementary data, which suggests that rds02
has positive connections to $107$ reference profiles, and negative
connections to $119$ profiles. However once again we know these
must be all false connections. Conversely, the new method
presented in this paper gave results that agree with the
underlying truth. With the criteria set above there was 1
connection found significant ($p<1/453=0.0022$). Since we expect
on average 1 false connection, this declared significant
connection can be taken as false.

The two examples of random gene signatures above revealed that on
the individual treatment instance level Lamb et al's Connectivity Map
gave many false connections because no safeguard was provided for
these. The reason for this was in the way the connectivity scores
were calculated. Briefly, Lamb et al's connectivity score was based on
the Kolmogorov-Smirnov statistic. At first two K-S statistics were
calculated, one for the up-regulated genes, and one for the
down-regulated, then these two values were combined to give a
single connectivity score. This way of defining the connection
score was unnecessarily complex, which made the calculation of a
possible p-value on an individual level difficult, if not
impossible.

\begin{table}
\caption{ \label{tab-rds01-sdz} The connection scores for the
random gene signature rds01 using the new methods.
To control the expected number of false connections at 1, the
threshold p value was set at $1/453=0.0022$. No connection was
found to be significant, which agrees with the underlying truth.}

\begin{tabular}{clccc}
\hline
rank&Compound&ID&score&pvalue\\
\hline
1&(-)-catechin&1101&0.334&0.003\\
2&sirolimus&1022&0.312&0.006\\
3&phentolamine&1138&0.309&0.007\\
4&5162773&892&0.292&0.011\\
5&resveratrol&595&0.286&0.013\\
$\cdots$&$\cdots$&$\cdots$&$\cdots$\\
451&felodipine&848&-0.002&0.988\\
452&estradiol&988&-0.001&0.991\\
453&bucladesine&959&0.001&0.993\\
\hline
\end{tabular}
\end{table}

On the treatment set level, Lamb et al's Connectivity Map provides a
permutation p value when a set of treatment instances associated
with the same compound were viewed as a whole. The permutation p
value was calculated by comparing a statistic (again based on K-S
statistic) of the treatment set with the distribution of many
random-set statistics. Those random sets were formed by randomly
selecting treatment instances from all instances. From the way the
permutation p value was calculated, it was clear that the
following null hypothesis was being tested: {\em The set of
instances in question have the same pattern (distribution) of
connections with the query gene signature compared with a randomly
formed set of the same size.} A null hypothesis such as this is
not the most appropriate one to test, because it is not directly
relevant to the question of whether the query gene signature had
any real biological connection to the treatment set. Lamb et al themselves also recognized that more rigorous methods for the estimation of statistical significance were probably required\citep{Lamb-etal-Science-2006}.
Putting its appropriateness
aside, if we do use the permutation p values from Lamb et al's Connectivity Map to control false connections, the same criteria
for setting threshold p value can be used. The results of
significance analysis on the connections between the two random
gene signatures and the $164$ treatments sets can be found in the
supplementary data. In this instance both connectivity maps gave
the right answers for these two random gene signatures. However,
for Lamb et al's Connectivity Map permutation p values were not
available for all the $164$ treatment sets. For those treatment
sets which only contain $1$ treatment instance or those sets with
mean connectivity score $0$, no permutation p values could be
calculated, and hence no statistical significance. This problem
affects the coverage and consequently the sensitivity of Lamb et al's Connectivity Map, because real biological connections between a
query gene signature and any of those treatment sets may not be
recognized.

\subsection{Querying the connectivity maps with experimentally derived real gene
signatures}

\begin{figure}
\centerline{\includegraphics[width=8.5cm,angle=0]{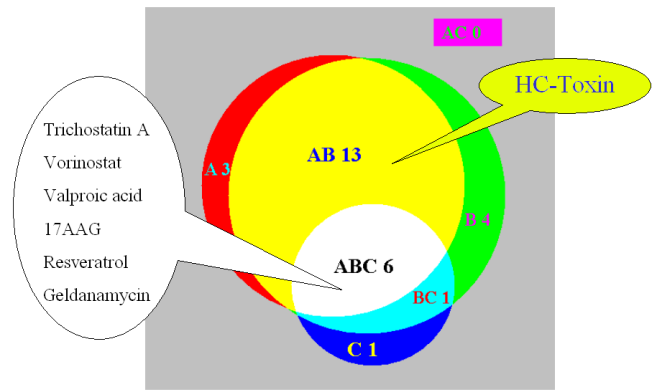}}
\caption{HDAC inhibitor gene signature: The Venn diagram
summarizing the findings of significant connections as identified
by the connectivity maps. A represents the instance-level analysis
using the new method; B, the set-level analysis using the new
method; C, the set-level analysis using Lamb et al's Connectivity Map.
The label ``AB 13'' means that $13$ compounds are identified as
significant solely by A and B (not C), ``B 4'' means that $4$
compounds are identified as significant solely by B (not A, not
C), and so on. The areas are approximately proportional to the
numbers they represent.} \label{fig-HDAC}
\end{figure}

\subsubsection{HDAC inhibitors}
To test the ability of the new method for identifying real
biological connections we utilized some of the same examples used
in \citep{Lamb-etal-Science-2006} to compare with Lamb et al's Connectivity Map. The first example was a gene signature of
histone deacetylase (HDAC) inhibitors (Lamb's gene signature
sigs01), which was compiled from
 \citep{Glaser-etal-MolCancerTher-2003} on the responses of T24,
MDA435 and MDA468 cells respectively to three histone deacetylase
(HDAC) inhibitors: vorinostat, MS-27-275, and trichostatin A. This
gene signature consisted of 8 up- and 5 down-regulated genes,
represented by $25$ Affymetrix probe-set IDs on the Affymetrix
HG-U133A microarrays.


As Lamb et al's Connectivity Map does not provide any safeguard on
individual treatment instance level against possible false
connections, we can only use the results of Lamb et al's Connectivity Map on the treatment set level. In total 6 compounds (vorinostat,
trichostatin A, resveratrol, geldanamycin, valproic acid, and
17AAG) were identified to have significant positive connectivity
with the HDAC inhibitor gene signature; and 2 compounds (5182598
and fludrocortisone) had significant negative connectivity.
Vorinostat, trichostatin A, and valproic acid are known HDAC
inhibitors thus the identification of these can be regarded as a
success. However another known HDAC inhibitor HC-toxin, a
reference profile of which was contained in the database, was not
identified. This happened because there was only one treatment
instance of the compound HC-toxin in the database and so no
permutation p value could be obtained using Lamb et al's method.
Based on their instance level results, Lamb et al highlighted HC-Toxin in \citep{Lamb-etal-Science-2006} as it had the 7th highest connectivity score (0.914) of all instances in the dataset. However, as the two examples of random gene signatures showed Lamb et al's instance level analysis gave false connections even for the highest connectivity scores 1.0. So the rational choice is to disregard the instance level results from Lamb et al's Connectivity Map.

We used this same HDAC inhibitor gene signature to query the new
connectivity map presented here, with p values calculated at both
the individual instance level and the treatment set level. $56$
treatment instances, representing $22$ distinct compounds, were
found to have significant connections to sigs01. On the treatment
set level $24$ compounds were found to have significant
connections with the signature.  Near the top of the significant
connection list were those known HDAC inhibitors highlighted in
 \citep{Lamb-etal-Science-2006}. Importantly though also included
in the output was HC-Toxin, which was not identified by the set-level analysis Lamb et al's Connectivity Map.
The full results of querying both connectivity maps using
the HDAC inhibitor gene signature are included in the
supplementary data. In Fig. \ref{fig-HDAC}, we summarize the
number of significant connections as identified by: (A) Instance
level analysis using the new method presented here; (B) Set level
analysis using the new method; (C) Set level analysis using Lamb et al's Connectivity Map. In total, our method (A) and (B) combined
identified 27 compounds and Lamb et al's Connectivity Map identified 8
compounds, as having significant connections to the HDAC
inhibitors. Our method missed only 1 of the 8 compounds found
significant by Lamb et al's Connectivity Map, while the latter missed
20 of the 27 compounds identified by our method, with HC-toxin
among the 20 compounds that were missed. The HDAC inhibitors
example thus shows that our new method is more sensitive at
detecting real connections. With the increased sensitivity and  false connections being properly controlled, the potential benefit of our method is obvious. In this example, the set-level analysis of Lamb et al's Connectivity Map identified 8 compounds with a false discovery rate of  $12.5\%$ (1/8), while the set-level analysis using our method identified 24 compounds with a false discovery rate of $4.2\%$ (1/24). Based on the findings of the connectivity maps, researchers can prioritize a small sub set of those compounds for further investigations and/or developing new biological hypotheses. For this example, using Lamb et al's Connectivity Map, the chance of pursuing a false connection is $12.5\%$, while using our method it is much lower at $4.2\%$. In practice this would save the researchers an enormous amount of time and resources and increase their rate of success.

\subsubsection{Estrogens}

The second example was a gene signature (Lamb's gene signature
sigs02) taken from a study  \citep{Frasor-etal-CancerRes-2004} of
MCF7 cells treated with estradiol. This gene signature consisted
of $40$ up- and $89$ down-regulated genes represented by $189$
Affymetrix probe-set IDs on the Affymetrix HG-U133A microarrays.
We queried both Lamb et al's Connectivity Map and the connectivity map
presented in this paper with this estrogen signature.

For the same reason given above, we only used the treatment set
level results of Lamb et al's Connectivity Map, which identified 4
compounds (genistein, estradiol,  tretinoin,
 and alpha-estradiol) as having significant positive connectivity
with the estrogen signature; and 5 compounds (trichostatin A,
fulvestrant, LY-294002,  vorinostat, and geldanamycin) that had
significant negative connectivity.

With our new connectivity map set-level analysis $16$ compounds
were found to have significant positive connection, and $25$
compounds had significant negative connection to the estrogen gene
signature. The $16$ compounds with positive connection included
genistein, estradiol, and alpha-estradiol, all known to be
estrogen receptor agonists. The $25$ compounds with negative
connection included fulvestrant, raloxifene and tamoxifen, all
known to be estrogen receptor antagonists. In comparison, the Lamb
method identified all the estrogen receptor agonists above, but
missed all the estrogen receptor antagonists except fulvestrant.
These results therefore indicate the sensitivity of our new method
is substantially increased. The Lamb method was able to detect the
pure estrogen receptor antagonist fulvestrant, but missed the two
compounds tamoxifen and raloxifene which have mixed antagonist and
agonist estrogen receptor activities.

The results from our set level analysis also suggest that
nordihydroguaiaretic acid (NDGA) has significant positive
connection with estradiol. This connection is supported by recent
studies
 \citep{Fujimoto-etal-LifeSciences-2004,Sathyamoorthy-etal-CancerRes-1994},
where NDGA has been shown to have estrogenic activity and able to
elicit an estrogen-like response. Another compound monorden
(radicicol), suggested by our method as having negative connection
to the estrogen gene signature, has been shown to repress the
transcriptional function of the estrogen receptor
 \citep{Lee-etal-MolCelEndoc-2002} which suggests that it may have
some estrogen receptor antagonist-like properties. The tabulated
results of querying both connectivity maps using the estrogen gene
signature are included in the supplementary data.
Fig.\ref{fig-Estrogen} summarizes the numbers of significant
connections identified by both connectivity maps. All $9$
compounds found significant by Lamb et al's Connectivity Map were also
identified by the new method (either on the instance level or on
the set level or both). However many compounds identified as
significant with either positive or negative connection to
estradiol using our method were not identified by Lamb et al's Connectivity Map, included amongst these were raloxifene,
tamoxifen, monorden, and NDGA.

\begin{figure}
\centerline{\includegraphics[width=8.5cm,angle=0]{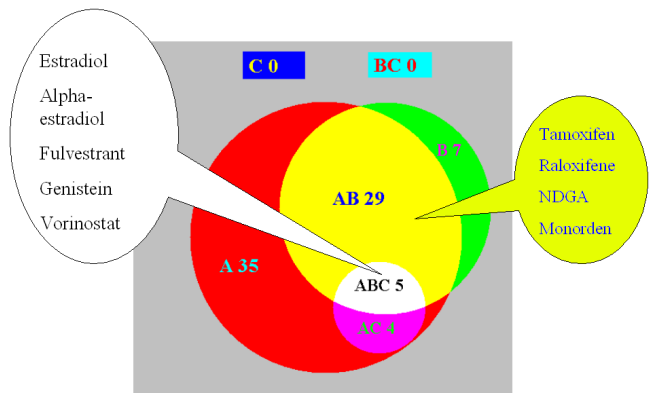}}
\caption{Estrogen gene signature: The Venn diagram summarizing the
findings of significant connections as identified by Lamb et al's Connectivity Map and the new method here. The labelling follows
the same conventions as in the previous figure.}
\label{fig-Estrogen}
\end{figure}

\subsubsection{Immunosuppressive therapy}
For further testing we compiled a new gene signature from a study
on cardiac allograft rejection and response to immunosuppressive
therapy \citep{Horwitz-etal-Circulation-2004}, where patients were
treated with standard immunosuppression with corticosteroids,
antimetabolites, calcineurin inhibitors, and/or sirolimus. This
gene signature consisted of 40 Affymetrix probe set IDs (see Table
2 of  \citep{Horwitz-etal-Circulation-2004}). Our set-level
analysis identified 29 compounds as having significant connections
with this gene signature. The three top compounds were
azathioprine, thalidomide, and rosiglitazone. Azathioprine is a
commonly used immunosuppressive drug
\citep{Armstrong-Oellerich-Clinical-Biochemistry-2001,Matalon-etal-ReproductiveToxicology-2004},
so its significant positive connection with the gene signature is
a good indication that our new connectivity map works very well
here. The second compound thalidomide, which had a positive
connection score, also has known immunosuppressive activities
\citep{McHugh-etal-ClinExpImmunol-1995}, inhibits release of
TNF$\alpha$ from monocytes, and modulates other cytokine actions.
The recognized properties of these molecules therefore accord with
the outcome of the connectivity matching. The third compound
rosiglitazone had a negative connection with the signature
suggesting it may have properties to reduce or mitigate the
effects of immunosuppressive activity. Recently, rosiglitazone was
reported to suppress cyclosporin-induced chronic transplant
dysfunction and prolong survival of rat cardiac allografts (Chen
et al, 2007, Transplantation)
\nocite{Chen-etal-Transplantation-2007}, where cyclosporin is also
a commonly used immunosuppressive drug
\citep{Warrington&Shaw-ExpertOpinion-2005}.

At the instance level, our method identified $89$ reference
profiles as having significant connections to the
immunosuppressive drug gene signature, representing $63$ distinct
compounds. The top 3 compounds were azathioprine, staurosporine,
and trichostatin A, which all achieved positive connection scores
with this gene signature. The second compound, staurosporine, a
protein kinase C inhibitor, is classified as an antineoplastic and
immunosuppressive antibiotic drug
\citep{Ting-etal-ImmunoPharm-1995}.
The third compound, trichostatin A, was recently shown to have
some immunosuppressive effects in leukemia T cells
\citep{Januchowski-Jagodzinski-InternationalImmunopharmacology-2007}.
Therefore the method of instance testing could be particularly
valuable for the identification of pharmacological and
toxicological properties in novel molecules.

\section{Discussion} \label{sec-discussion}
We have presented in this paper a new framework for a connectivity
map, with the advantages that statistical significance measures
are calculated at both treatment instance level and treatment set
level, thus providing effective control over false connections.
This important safeguard was not available in Lamb et al's Connectivity Map at the instance level, as revealed by the two
examples of random gene signatures.
As the connectivity maps are most useful for high throughput screening and for generating new biological hypothesis, it is crucial that false connections are tightly controlled, such that the users do not end up chasing false connections and wasting time and money.
We compared the performance of Lamb et al's Connectivity Map and the new method
presented in this paper, using two gene signatures compiled by Lamb et al
and also a new gene signature for immunosuppressive drugs. All
these examples demonstrated that our new method is more sensitive
and gives more robust results than Lamb et al's method.

The set-level analysis of Lamb et al's Connectivity Map tests an
implicit null hypothesis that is not most appropriate. This can be
seen more clearly from recent studies on the significance analysis
of gene sets
\citep{Tian-etal-PNAS-2005,Efron-Tibshirani-AnnApplStat-2007}.
Tian et al were among the first groups of authors who made
explicit distinctions between two null hypotheses
 \citep{Tian-etal-PNAS-2005} concerning a set of entities ( a set
of genes in context of Tian et al's paper). In the present context
of connectivity map, the two null hypotheses are:

1. Hypothesis $Q_1$: The treatment instances in a set show the
same pattern of connections with the query gene signature compared
with the rest of the treatment instances.

2. Hypothesis $Q_2$: The treatment set does not contain any
treatment instances which have real connections with the query
gene signature.

Tian et al's discussions over the relationship between $Q_1$ and
$Q_2$  \citep{Tian-etal-PNAS-2005} apply to a treatment set as
follows: Given all the $453$ treatment instances, even if none of
them have real connection to the query gene signature, the
observed connection scores of a treatment set could still be very
different from those treatment instances outside of the set
because of the special correlation structure among the treatment
instances within the set. Chen et al also raised some concerns
with testing the hypothesis $Q_1$ (Chen et al, 2007,
Bioinformatics)\nocite{Chen-etal-Bioinformatics-2007}, that this
hypothesis does not test whether the treatment set has
above-random connections to the query gene signature, but rather
it tests whether the observed connections in the set are more or
less than a randomly formed set of same size. We agree with Chen
et al's assessment on the inappropriateness of $Q_1$. It was clear
that Lamb et al's Connectivity Map was testing the null hypothesis
$Q_1$. Here we tested $Q_2$ which is more appropriate.

Our results indicate that the method presented here can identify
many significant connections to a query gene signature. Then what
criteria should we use and which compounds should we choose if new
biological hypotheses are to be developed? Our suggestion is to
concentrate more on those compounds which have many replicate instances in
the connectivity map. Because the results obtained for those big
treatment sets do not depend heavily on the quality of a small
number of treatment instances, as in the case of small treatment
sets or singleton sets (treatment sets with only 1 instance each).
Lamb et al also recognized the importance of having replicate instances, and noted that the power to detect connections might be greater for compounds with many replicates.
In defining a treatment set, ideally only the treatment instances of the same compound with the same dose and the same cell type should be considered as a set. For example, the biological state of  HL60 cells perturbed by raloxifene should be considered as a different biological state from that of MCF7 cells perturbed with the same raloxifene, thus these two instances of raloxifene should not be put into the same set.  In this paper for comparison purpose we adopted Lamb et al's choice in defining a treatment set, i.e., all the instances of a compounds were grouped together as a set regardless of their possible differences in dose or cell type.
Mixing the instances of a compound with different doses and/or different cell types increases the heterogeneity of an otherwise more homogenous treatment set. This tends to average out the distinct characteristics attributable to the cell type or dosage difference, making some set-level connections insignificant or their interpretation difficult. In such cases the instance level connections supported by statistical significance can be of great help in interpreting the results. For future connectivity maps, efforts should be made to provide as many more replicate treatment instances (replicates with the same compound, the same dose, and the same cell type, etc) so that the undesirable reliance on individual instances can be minimized.


\section{Acknowledgments}
\noindent We wish to acknowledge the support of the microarray
team of the MRC Toxicology Unit. SDZ thanks Qing Wen for helpful
discussions on a searching algorithm in the implementation of the
new connectivity map.


\end{document}